# Community Detection and Growth Potential Prediction from Patent Citation Networks

Asahi Hentona[1]
s173348@stn.nagaokaut.ac.jp

Hirofumi Nonaka[1]
nonaka@kjs.nagaokaut.ac.jp

Kensei Nakai[1]
s163405@stn.nagaokaut.ac.jp

Takeshi Sakumoto[1]
s183353@stn.nagaokaut.ac.jp

Shotaro Kataoka[1]
s183347@stn.nagaokaut.ac.jp

Elisa Claire Alemán Carreón[1]
s153400@stn.nagaokaut.ac.jp

Hugo Alberto Mendoza España[1]
s173330@stn.nagaoka.ac.jp

Toru Hiraoka[2]
hiraoka@sun.ac.jp

Masaharu Hirota[3]
hirota@mis.ous.ac.jp

[1]Department of Information and Management Systems Engineering, Nagaoka University of Technology, Niigata, Japan

[2]Department of Information Systems, University of Nagasaki, Nagasaki, Japan

[3]Department of Informatics, Okayama University of Science, Okayama, Japan

## ABSTRACT

The scoring of patents is useful for technology management analysis. Therefore, a necessity of developing citation network clustering and prediction of future citations for practical patent scoring arises. In this paper, we propose a community detection method using the Node2vec. And in order to analyze growth potential we compare three "time series analysis methods", the Long Short-Term Memory (LSTM), ARIMA model, and Hawkes Process. The results of our experiments, we could find common technical points from those clusters by Node2vec. Furthermore, we found that the prediction accuracy of the ARIMA model was higher than that of other models.

## CCS CONCEPTS

• Computing methodologies~Machine learning • Mathematics of computing~Probability and statistics

## KEYWORDS

Patent analysis, Community detection, Growth prediction, Node2vec, ARIMA, LSTM, Hawkes process

## 1 INTRODUCTION

Patents can protect a firm's inventions from others' appropriation and bring about economic benefits [1]. As such, patent data is used for a product's developmental status, market competition and other important aspects to analyze technology management. As shown in previous research, it can be used to identify trends in the industry as well as the competitive power of enterprises or countries [2,3], to quantize R&D and innovative activities [4]. Some researchers employed a text-mining approach to analyze the contents of patent documents. Uchida et al. [5] developed a method for automatically generating patent maps (not technology-effect patent maps) using a concept-based vector space model. Ishikawa et al. [6] tried to extract technology terms from a patent document using the clue phrase "because of". A clue phrase was defined as the word linking a technology term and an effect term. Sakai et al. [7] proposed an automatic collection of clue phrases from patent documents by using an entropy-based score, and Nonaka et al. [8] expanded it to extract information on the effect and the technological terms from patent documents by using grammar patterns.

While information extraction of patent documents is important to analyze technology, the scoring of patents is also useful for technology management analysis. There are many aspects of patent scoring such as patent application number-based index [9]. Among them, the number of cited patents is specifically used for quantification of patent value as it shows the degree of attention from other competitors in the same technology field. Hall [10] found that patent citation significantly affects market value. Chang's study [11] employs a panel threshold regression model to test whether the patent i-index has a threshold effect on the relationship between patent citations and market value in the pharmaceutical industry. However, simple citation counting, or its expansion, should not be considered the inherent value of importance of different patents. Therefore, it is necessary to assign importance ranks to each patent. To address the issue, a link evaluation method such as PageRank [12] or HITS [13] has been used for calculating importance of patents. Lukach, et al. [14] have proposed computing importance by the PageRank score of patents. According to their study, PageRank patent importance weights differs from the weights based on the number of backward or forward citations. Nonaka, et al. [15] calculated a correlation between the patent score based on HITS, which is similar with Pagerank score, and stock data. They indicated high correlation between the patent score of B to B companies and their time series of stock price value. Additionally, Bruck [16] showed that using the PageRank score of patent documents gives the possibility to make predictions about future technological development. While link evaluation methods are useful for patent scoring, these methods are based on static models and thus do not reflect the growth potential of the technology cluster of the patent. Therefore, even if the cluster of a patent has no hope of growing, we recognize the patent is important if PageRank or other ranking score is high. Therefore, there arises a necessity of developing citation network clustering and prediction of future citations. In this paper, we detect a community using Node2vec [17] which shows high performance of network embedding. And, to evaluate growth potential, we compare performance of 3 models; the Long Short-Term Memory (LSTM) [18] deep learning model, Autoregressive Integrated Moving Average (ARIMA) [19], and Hawkes process [20]. Node2vec is a novel network embedding model using random walk and Skip-Gram to group similar nodes to communities. Long Short Time Memory deep learning model is widely used for time series data such as financial data [21]. ARIMA model is a traditional time series forecasting method and applied in various fields such as economics [22]. Hawkes process is a self-exciting process, it has been applied to model highly clustered event sequences, such as seismology [23].

# 2 METHODOLOGY

## 2.1 Overview of our method

Our proposed method is shown in Figure 1. First, we extracted the citation network from the patent database. Then we detected patent communities in the citation network. Finally, to predict growth potential of each community we compare the performance of three models, LSTM, ARIMA, and Hawkes process.

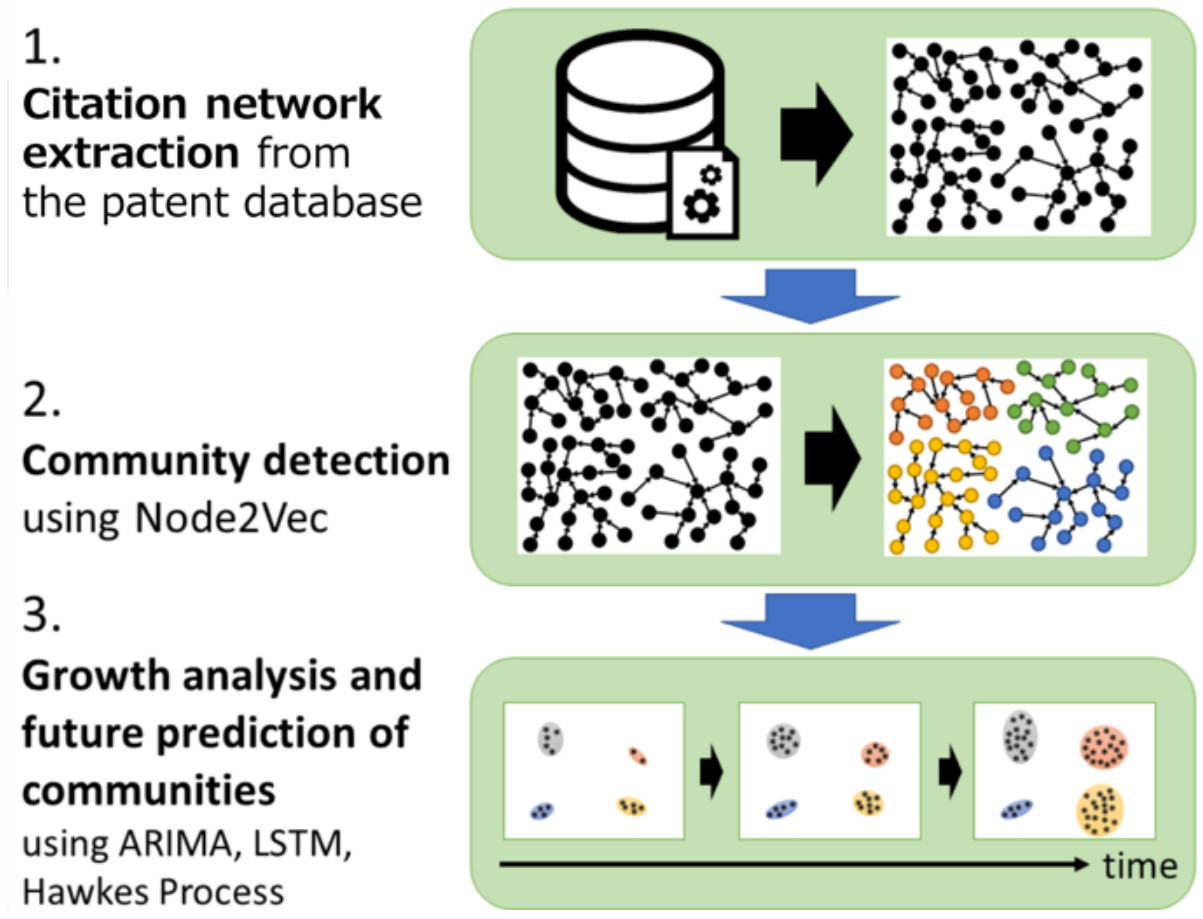

**Figure 1: Outline of our method**

## 2.2 Citation network graph extraction from patent database

Using the patent information database as an input, we created a graph file of the patent citation network in the patent field to be analyzed. First, we prepared the patent information database. In this study, we used a Japanese patent database provided by the Institute of Intellectual Property of Japan (IIP) [24]. It contains about 6,000,000 patents from 1960 to 2017 and their examiner citations. After that, we extracted the patent citations network in the patent field to be analyzed from the database and generates a graph file (DOT file). In the graph file, the nodes of the network (patent application number) and the link (quotation relation of the patent) are recorded.

## 2.3 Community detection using Node2vec

To detect a community from large scale networks, mapping each node of the network into low dimensional vector is crucial because the vector is useful for clustering networks. In this research, we used Node2vec, which is like Word2vec that is often utilized by Skip-gram [25], to address the issue. Skip-gram is based on the distributional hypothesis that words in similar contexts tend to have similar meanings [26], in other words, similar words tend to appear in similar word neighborhoods. In this model, the word feature representations are learned by optimizing the likelihood objective using SGD with negative sampling [27] which effectively reduce computational cost. To enhance the Skip-gram model towards network information, recent researches such as DeepWalk [28] and Node2vec have used an analogy between a network and a "document" which is an ordered sequence of words. These researches sample sequences of nodes randomly from the network and turn the network into an ordered sequence of nodes and documents.

DeepWalk has been widely used for learning the representations of nodes in a network, which can be extracted from features of the neighbor structures of nodes. DeepWalk discovers that the distribution of nodes appearing in short random walks is similar to the distribution of words in natural language. DeepWalk used a stream of short random walks on a network to generate a set of walk sequences as well as short sentences and phrases in the case of a "document". For each walk sequence $s = \{v_1, v_2, \cdots, v_s\}$, following Skip-Gram, DeepWalk aims to maximize the probability of the neighbors of node vi in this walk sequence as follows:

$$Pr\left(\{v_{i-w}, \cdots, v_{i+w}\} \setminus v_i \mid \Phi(v_i)\right) = \prod_{j=i-w, j\neq i}^{i+w} Pr\left(v_j \mid \Phi(v_i)\right) \quad (1)$$

where $w$ is the window size, $\Phi(v_i)$ is the current representation of $v_i$ and $\{v_i - w, \cdots, v_i + w\} \setminus v_i$ is the local context nodes of $v_i$. Finally, DeepWalk uses hierarchical soft-max to infer the probability of its neighbors in the walk.

Node2vec improves the performance of DeepWalk to design a second order random walk strategy to sample the neighborhood nodes, which can smoothly interpolate between breadth-first sampling (BFS) and depth-first sampling (DFS). To adopt BFS, the neighborhood is restricted to nodes which are immediate neighbors of the source. On the other hand, the neighborhood consists of nodes sequentially sampled at increasing distances from the source node using DFS. Real-world networks commonly show a mixture of both attributes. To mix the two sampling strategies adequately, Node2vec uses two parameters $p$ and $q$ which guide the walk. Considering a random walk that just traversed edge $(t, v)$ and now resides at node $v$. The unnormalized transition probabilities $\pi_{vx}$ on edges $(v, x)$ leading from $v$ is the following.

$$\pi_{vx} = \alpha_{pq}(t, x) = \begin{cases} \frac{1}{p} & \text{if } d_{tx} = 0 \\ 1 & \text{if } d_{tx} = 1 \\ \frac{1}{q} & \text{if } d_{tx} = 2 \end{cases} \quad (2)$$

where $d_{tx}$ denotes the shortest path distance between nodes $t$ and $x$, and $\alpha_{pq}(t, x)$ denotes the unnormalized transition probabilities when weighted graphs. In case of the unweighted graphs (i.e. the weight is 1), $\pi_{vx} = \alpha_{pq}(t, x)$ .

Parameters p and q can be controlled to interpolate between BFS and DFS, and, thereby, reflect an affinity for different notions of node equivalences. Parameter $p$ controls the likelihood of immediately revisiting a node in the walk. Setting it to a high value ($> max(q, 1)$) ensures that we are less likely to sample an already visited node in the following two steps (unless the next node in the walk had no other neighbor). This strategy encourages moderate exploration and avoids 2-hop redundancy in sampling. On the other hand, if $p$ is low ($< min(q, 1)$), it would lead the walk to backtrack a step (Figure 2) and this would keep the walk "local" close to the starting node $u$. Parameter $q$ allows the search to differentiate between "inward" and "outward" nodes. If $q > 1$, the random walk is biased towards nodes close to node $t$. Such walks obtain a local view of the underlying graph with respect to the start node in the walk and approximate BFS behavior in the sense that our samples comprise of nodes within a small locality. In contrast, if $q < 1$, the walk is more inclined to visit nodes which are further away from the node $t$. Such behavior is reflective of DFS which encourages outward exploration. However, an essential difference here is that we achieve DFS-like exploration within the random walk framework. Hence, the sampled nodes are not at strictly increasing distances from a given source node $u$, but in turn, we benefit from tractable preprocessing and superior sampling efficiency of random walks. Note that by setting $\pi_{vx}$ to be a function of the preceding node in the walk $t$, the random walks are 2nd order Markovian.

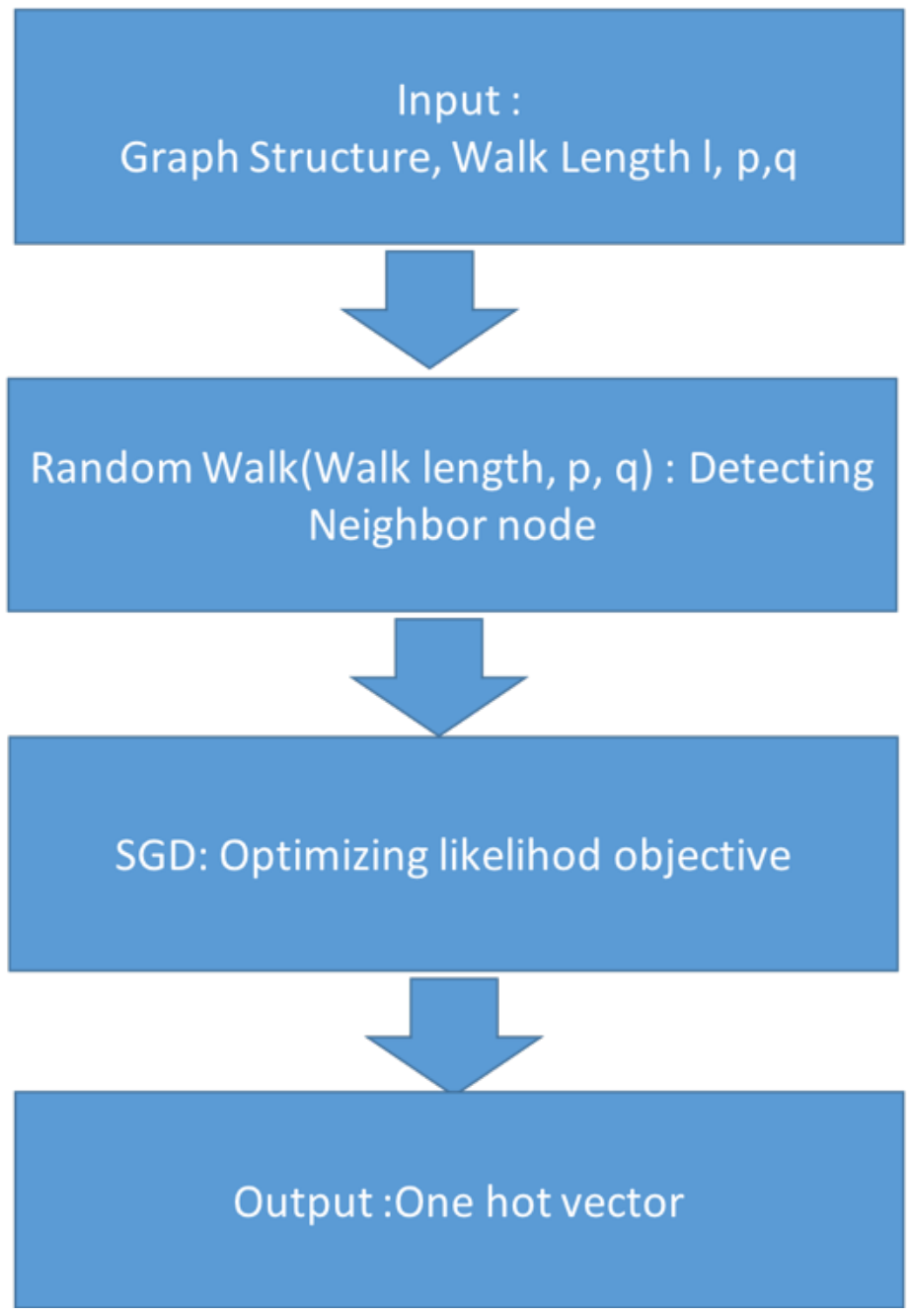

**Figure 2: Overview of the Node2vec algorithm**

## 2.4 Growth analysis and prediction of communities using Hawkes process

In this research, we evaluated the performance of Hawkes process to predict growth potential of a patent community.

Point processes are collections of random points falling in some space, such as time and location. The simplest class of point process is the Poisson process. Events following the Poisson process are arriving at an average rate of event intensity per unit time. Hawkes processes, which is a self-exciting stochastic process and an extension of pure Poisson process, models a sequence of arrivals of some type over time. Self-excitation means occurrence of an event causes other similar events to tend to occur. Since Hawkes process is a self-exciting process, it has been applied to model highly clustered event sequences, such as seismology [29] and their offspring events [30] or credit default events, high frequency trading in finance market [31-33]. The clustered events are caused by interconnections between entities or the potential to connect entities inside a network. Therefore, Hawkes Process have been also used to Patent citation forecasting [34].

The intensity of such self-excitation depends on the historical path of the original process. Now we consider a counting process $N_t$, which is a stochastic process $(N_t : t \geq 0)$ taking values in $N_0$ that satisfies $N_0 = 0$, is almost surely finite and is a right-continuous step function with increments of size +1, jumps up by one when an event arrives, and describes the number of arrivals in the time interval $[0, t]$. Here, the intensity of a counting process is defined as the conditional expected arrival rate during a short time given the historical path of all events up to the current time. The formulation is the following,

$$\lambda_t = \lim_{\Delta t \to 0} \frac{E[N_{t+\Delta t} - N_t | H_t]}{\Delta t} \quad (3)$$

where $H_t$ is the history up until the last arrival and following the counting process $N_t$ up to time $t$. For a pure Poisson process, the intensity $\lambda$ is given as a constant. On the other hands, Hawkes processes have the following stochastic intensity and is a nonhomogeneous Poisson process $N_t$ with an intensity $\lambda_t$ described by a constant base intensity $\mu$ and a self-exciting term that is a weighted sum over previous events.

$$\lambda_t = \mu + \int_{-\infty}^{t} \Phi(t-u) dN_u \quad (4)$$

The "basis constant" $\mu$ is a nonnegative value and the "influence kernel" $\Phi(\tau) \geq 0$ describes the effect on the instantaneous event rate $\lambda_t$ of a past event that took place at time $t - \tau$. The parameter $\mu$ is the intensity if there has been no past arrival of any jumps, and the function $\Phi$ is referred to as the generator of the Hawkes process that governs the density of arrival rate of events. Note that if a generator $\Phi$ is zero, Hawkes processes are simplified to pure Poisson processes.

According to Hyun [34], expectation of $N_t$, in other words, prediction value of $N_t$, is achieved by computing the first moments of the Hawkes process $N_t$. Using the stochastic integration on the Hawkes process and their intensity process, the closed formulas of first and second moments are obtained as (5) and (6), respectively:

$$E[N_t] = \left(\frac{\lambda_0}{\alpha-\beta} + \frac{\beta\mu}{(\alpha-\beta)^2}\right) e^{(\alpha-\beta)t} - \left(\frac{\beta\mu}{\alpha-\beta}\right) t - \left(\frac{\lambda_0}{\alpha-\beta} + \frac{\beta\mu}{(\alpha-\beta)^2}\right) \quad (5)$$

$$E[N_t^2] = \frac{B}{2(\alpha-\beta)} e^{2(\alpha-\beta)t} - \frac{2(A+B-C)}{\alpha-\beta} e^{(\alpha-\beta)t} - C(\alpha-\beta)t^2 + (\lambda_0 + 2A + 2C)t + \frac{4A+3B-4C}{2(\alpha-\beta)} \quad (6)$$

$$where\ A = \frac{\lambda_0(\alpha^2 + 2\beta\mu) - 2\lambda_0\alpha(\alpha-\beta)}{2(\alpha-\beta)^2},$$

$$B = \frac{\lambda_0(\alpha^2 + 2\beta\mu) + 2\lambda_0\alpha(\alpha-\beta)}{2(\alpha-\beta)^2}\ and\ C = \frac{\lambda_0\beta\mu}{(\alpha-\beta)^2}$$

For the patent citation modelling, consider $N_t$ as being the number of times a patent has been cited until time $t$ since the birth of the patent. The parameter means the base intensity for which a patent has never been cited before, and the generator controls the jump size of the patent citation intensity when a new citation occurs.

To note is that Hawkes process cannot use data, which include events with the same arrival time. Thus, in this paper, when multiple patents are filed at time $t$, we have shifted the application time of each patent by a time that divided $N_t$ by unit time.

## 2.5 Growth analysis and prediction of communities using ARIMA

One of the most important and widely used time series models is the autoregressive integrated moving average (ARIMA) model Introduced by Box and Jenkins. ARIMA model is a traditional time series forecasting method and applied in various fields [35-36]. In an ARIMA model, the future value of a variable is supposed to be a linear combination of past values and past errors, expressed as follows

$$y_t = \theta_0 + \phi_1 y_{t-1} + \phi_2 y_{t-2} + \cdots + \phi_p y_{t-p} + \epsilon_t - \theta_1 \epsilon_{t-1} - \theta_2 \epsilon_{t-2} - \cdots - \theta_q \epsilon_{t-q} \quad (7)$$

where $y_t$ is the actual value and $\epsilon_t$ is the random error at time $t$, and $\theta_i$ are the coefficients, $p$ and $q$ are integers that are often referred to as orders of autoregressive and moving average polynomials, respectively. Random errors are assumed to be independently and identically distributed with a mean of zero and a constant variance of $\sigma^2$. (7) entails several important special cases of the ARIMA family of models. If $q = 0$, then (7) becomes an AR model of order $p$. When $p = 0$, the model reduces to an MA model of order $q$. One central task of the ARIMA model building is to determine the appropriate model order $(p, q)$.

Basically, this method has three phases: model identification, parameter estimation and diagnostic checking.

In the identification step, it is necessary to make the time series stationary. Therefore, if the observed time series is shown non-stationary, differencing and power transformation are often applied to the data to remove the trend and stabilize the variance. The parameters are estimated such that an overall measure of errors is minimized at parameter estimation step.

## 2.6 Growth analysis and prediction of communities using LSTM

In this research, we used long short-term memory (LSTM) networks, one of the most advanced deep learning architectures for sequence learning tasks, such as handwriting recognition, speech recognition, or time series prediction [37,38] to predict citation number of each community.

The model is described as following:

$$i_t = \sigma(W_{ix}x_t + W_{im}m_{t-1} + b_i) \quad (8)$$
$$f_t = \sigma(W_{fx}x_t + W_{fm}m_{t-1} + b_f) \quad (9)$$
$$c_t = f_t \odot c_{t-1} + i_t \odot g(W_{cx}x_t + W_{cm}m_{t-1} + b_c) \quad (10)$$
$$o_t = \sigma(W_{ox}x_t + W_{om}m_{t-1} + b_o) \quad (11)$$
$$m_t = o_t \odot h(c_t) \quad (12)$$
$$y_t = \phi(W_{ym}m_t + b_y) \quad (13)$$

where the $W$ terms denote weight matrices (e.g. $W_{ix}$ is the matrix of weis from the input gate to the input), the $b$ terms denote bias vectors (e.g. $b_i$ is the input gate bias vector), $\sigma$ is the hard sigmoid function, and $i, f, o$ and $c$ are respectively the input gate, forget gate, output gate and cell activation vectors, all of which are the same size as the cell output activation vector $m$, is the element-wise product of the vectors, $g$ and $h$ are the cell input and cell output activation functions, and $\phi$ is the network output activation function.

# 3 EXPERIMENTS AND RESULTS

## 3.1 Node2vec

We applied Node2vec to real patent networks to evaluate performance of our method. In this evaluation, we analyzed the citation network of the game technology field including "Roulette-like ball games and related technology fields" (IPC code: A63F5). Table 1 shows the details of the data. This experiment has been carried out, where the parameters of Node2vec are $p = 1, q = 0.5$ and $Walk\ Length = 80$.

**Table 1: Network to Be Analyzed**

| Patent classification name | Number of patent nodes | Number of cited links |
|---|---|---|
| Roulette-like ball games | 29621 | 74529 |

And we also applied hierarchical clustering (Ward's method) to the vectors generated by Node2vec in order to detect patent community. In this experiment, 20 % of the maximum value of the inconsistency values was set as the threshold value. Figure 3 shows the hierarchical clustering result. This figure is a graphical representation of the result of the hierarchical clustering. The cluster is integrated hierarchically inside the figure as the bottom layer clusters (number of clusters $K$ = *target patent number*) on the outside of the figure, and it finally shows that it connects to one cluster. As a result, 39 clusters were obtained from the citation network of Roulette-like ball games and related technology fields.

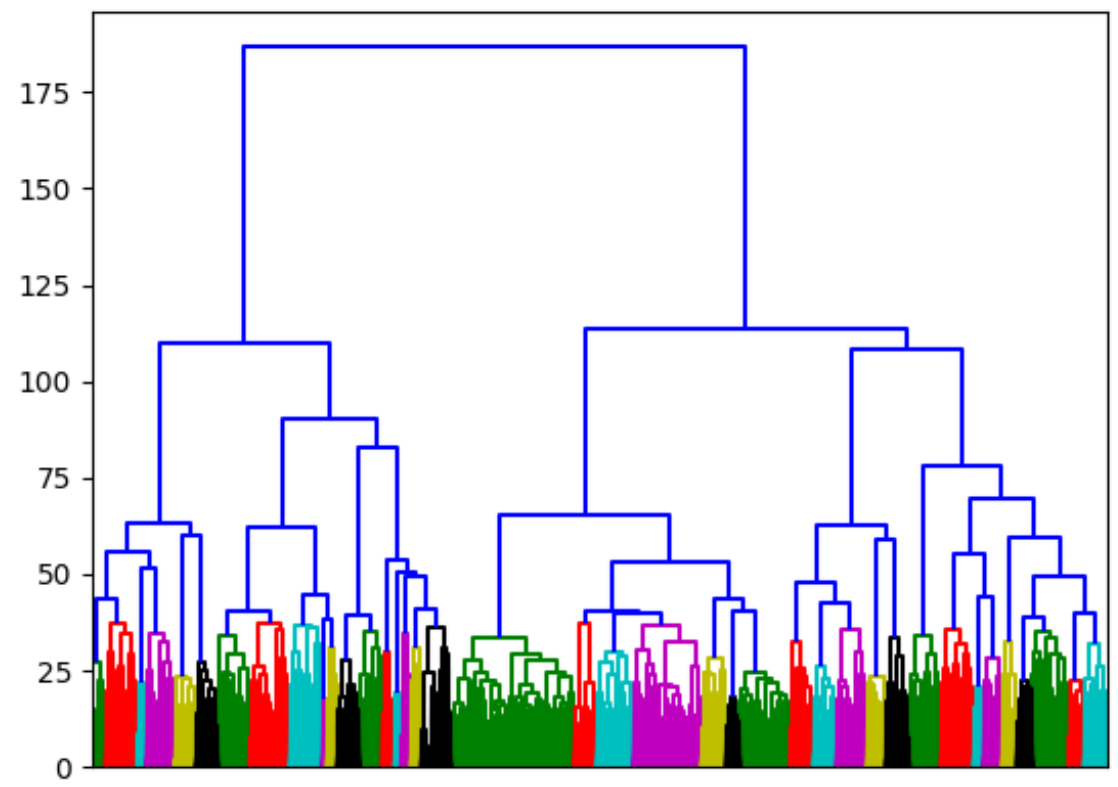

**Figure 3: A dendrogram of the results of running the vectors of Node2vec through hierarchical clustering. The X-axis indicates patents, and the Y-axis is the inconsistency values.**

## 3.2 Prediction models

We predicted the number of citations of each community to evaluate the growth potential of the community by ARIMA, LSTM, and Hawkes process. The unit time of the patent applications time series was set to a month, and the data of patents applications from 1985 to 2005 were used as a learning data. We used three models and this data, and we predicted a number of patent applications a quarter and a year ahead from December 2005.

Here, we describe setting parameters of each model. The optimum parameters $\alpha, \beta$ and $\mu$ of Hawkes process model were estimated by the Nelder–Mead method. The optimal parameters $p, q,$ and $d$ of ARIMA model searchingly calculated and obtained the parameters of a model that minimizes Akaike's Information Criterion. Next, we describe the parameters of LSTM. We have adopted that the cell input and cell output activation functions is ReLU, the network output activation function is linear function, the number of units is 128, the batch size is 32, the dropout rate is 0.2, the learning rate is 0.001, and the optimization algorithm is Adam. In addition, when the validation loss deteriorated 3 consecutive times, the learning rate was multiplied by 0.1, and when it deteriorated 10 consecutive times it was set to round up the learning.

To measure the effectiveness of those models, we used Mean Absolute Percentage Error (MAPE) and Direction Accuracy. MAPE is given by $\frac{1}{N}\Sigma_{i=1}^{K}\left|\frac{P_i(t)-R_i(t)}{R_i(t)}\right|$ where $P_i(t)$ denote predicted number of patent applications and $R_i(t)$ denote the real number of patent applications of cluster $i$ at time $t$. Direction accuracy is an evaluation index focusing only on whether the predicted value is higher or lower from the current point (in this experiment, it is December 2005). Direction Accuracy is given by $\frac{1}{N}\Sigma_{i=1}^{K}\frac{PD_i(t)}{PD_i(t)+RD_i(t)}$ where $PD_i(t)$ denote predicted direction and $RD_i(t)$ denote the real direction of cluster $i$ at time $t$.

Table 2 shows the prediction performance of Hawkes process, ARIMA, and LSTM, and Table 3 shows the number of clusters and Direction accuracy within only clusters that absolute error of predicted value is less than MAPE.

**Table 2: Performance of Models (All clusters)**

| | Model | MAPE [%] | Direction Accuracy [%] |
|---|---|---|---|
| One quarter | Hawkes | 95.22 | 43.59 |
| | ARIMA | 56.98 | 82.05 |
| | LSTM | 59.63 | 71.79 |
| One year | Hawkes | 88.48 | 58.97 |
| | ARIMA | 71.55 | 82.05 |
| | LSTM | 82.09 | 64.10 |

**Table 3: Performance of Models (Only clusters that absolute error of predicted value is less than MAPE)**

| | Model | Number of Clusters of Over MAPE | Direction Accuracy [%] |
|---|---|---|---|
| One quarter | Hawkes | 22 / 39 | 22.73 |
| | ARIMA | 14 / 39 | 71.43 |
| | LSTM | 16 / 39 | 62.50 |
| One year | Hawkes | 22 / 39 | 45.45 |
| | ARIMA | 7 / 39 | 85.71 |
| | LSTM | 5 / 39 | 40.00 |

# 4 DISCUSSION

## 4.1 Node2vec

Using the results of Node2vec, we examined the technical field by randomly referring to patent documents of patents belonging to each cluster. As a result, we could find common technical points.

In patent classification "ball game like roulette", there are 39 clusters. These clusters have been grouped adequately.

For example, a cluster is a collection of patents related to prevention of tampering with pachinko / pachislot machines and their peripheral technology. One patent of the cluster is a circuit board box (JP1994-269539) which makes it difficult to tamper with the slot machine. This patent is quoted for slot machines equipped with the circuit board box and patents applying the technology. Besides this, as a patent of this cluster, there is a structure for facilitating the inspection of pachinko machines, and a fraud detection mechanism for pachinko machines. For another example, a cluster is a set of patents related to the production technique of the game machine. The core patent is a slot display technique (JP 2004-357878). This patent is cited in a technical patent for realizing this method. In addition to this, the cluster has the structure of slot machine parts and the pachinko performance program.

In the above examination, we confirmed Node2vec is appropriate for graph clustering of patent citation networks.

## 4.2 Prediction models

ARIMA model had a higher accuracy than Hawkes Process and LSTM model in both MAPE and Direction Accuracy. In addition, it shows that even if the number of future patent applications are overestimated, it is possible to forecast the direction of whether a technology cluster grows or declines with an accuracy of 82.05 %.

The reason that the prediction accuracy of ARIMA model was high is presumed to be the possibility that the patent application series of each cluster is a unit root process. Accordingly, we used an Augmented Dickey-Fuller (ADF) test [39] to the patent application time series for each cluster. ADF test is a hypothesis test on whether time series data have unit roots. The more negative an ADF statistic is, the more likely it is to reject the hypothesis that there is a unit root at certain significance level $p$. We ran an ADF test under the condition that $p = 0.01$. As a result, it was found that the hypothesis regarding the presence of a unit root can't be rejected for the patent application time series of 13 clusters out of the 39 clusters. Since ARIMA model analyzes time series data after taking a difference of time series, linear trends are removed from the patent application series that are unit root processes. We concluded that the prediction accuracy of ARIMA model was higher than that of other models for this reason.

Although LSTM model is less accurate than ARIMA model, we consider that LSTM model unsuccessfully learned due to having a low amount of learning data. The length of the sequence data used in this experiment is 252, and the number of steps is set to 12. Because the learning data was few, we concluded that LSTM model did not fit.

Hawkes process model was the lowest accuracy than the other models in all predictions. In this research, we used the pure Hawkes process model in which the base intensity $\mu$ is a constant. We consider that low accuracy was caused by $\mu$ being a constant. Namely, used patent application series suggest that average value of the intensity fluctuates greatly with the lapse of time.

Table 2 shows that MAPE of Hawkes process model's forecast values for a year ahead is smaller than MAPE for a quarter ahead. Figure 4 shows that the predicted value of a fitted Hawkes process model tends to be estimated considerably lower than the actual measured value, and the actual measured value for a year ahead tends to be less than the one for a quarter. On the other hand, the average of the actual measured values for a quarter and a year are 8.67 and 5.31, respectively. It shows that the number of patent applications for one year ahead is smaller on average than the ones for a quarter ahead. Therefore, we concluded that ostensibly MAPE was getting smaller due to the number of applications after one year was less than that of the quarter. Hawkes process model is far from being strong in the long-term prediction than the short-term for this reason.

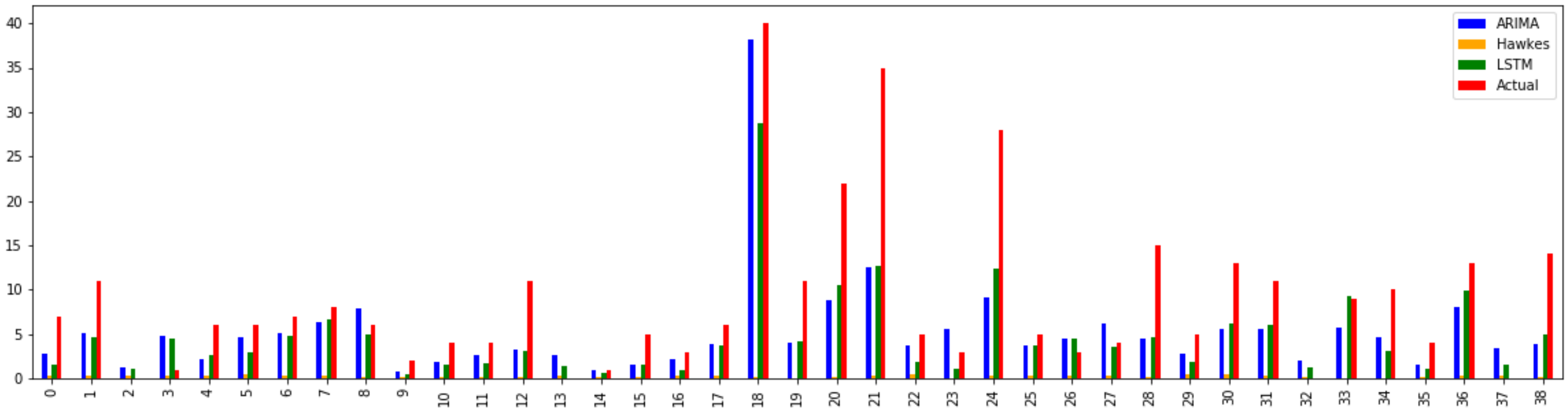

(a) Predicted values and the actual measured values a quarter after December 2005.

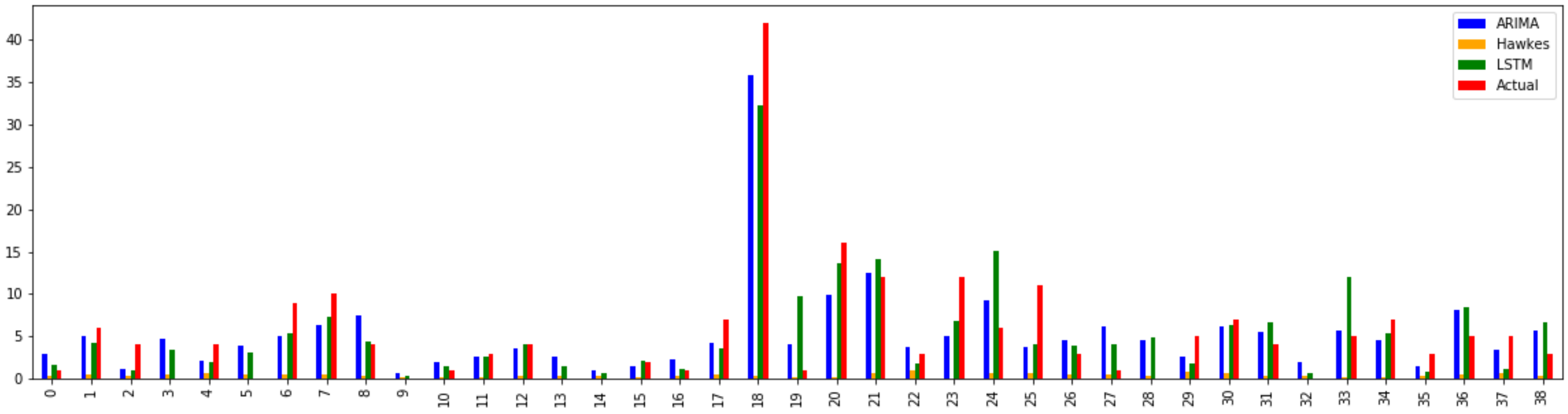

(b) Predicted values and the actual measured values a year after December 2005.

**Figure 4: Predicted values of each model and the actual measured values for each cluster. The Y-axis means the number of patent applications, and the X-axis means a cluster label.**

# 5 CONCLUSION

In this paper, the clustering of patent citation networks and the growth potential prediction were proposed with the aim of enabling corporate managers and investors to evaluate the scale and life cycle of technology.

In our proposed method, first, we extracted the citation network from the patent database. Then we detected patent communities in the citation network by Node2vec. Finally, to predict growth potential of each community we compared the performance of three models, Hawkes process, ARIMA, and LSTM. Then, using the results of Node2vec, we examined the technical field by randomly referring to patent documents of patents belonging to each cluster. As a result, 39 clusters were obtained from the citation network of Roulette-like ball games and related technology fields. We could find common technical points from those clusters, and the validity of Node2vec was shown. Finally, we predicted the number of citations of each community to evaluate the growth potential of the community by ARIMA, LSTM, and Hawkes process, and we compared the accuracy of the three models. We concluded that the prediction accuracy of ARIMA model was higher than that of other models.

In future work, we will apply this method to various technical fields. Moreover, we will conduct an experiment by using more data and verify the validity of our method.